\documentclass[nofootinbib,aps,prl,reprint,amsfonts,amssymb,amsmath]{revtex4-1}
\usepackage[bottom]{footmisc}
\usepackage{footnote}
\usepackage{perpage} 
\MakePerPage{footnote} 
\usepackage{hyperref}

\newcommand{\Z}{{\mathbb Z}}
\newcommand{\R}{{\mathbb R}}

\usepackage{color}
\usepackage{comment}
\usepackage{mathtools}
\begin{document}

\title{Four Dimensional $\mathbf{\mathcal{N}=4}$ SYM and the Swampland}

\author{Hee-Cheol Kim $^1$, Houri-Christina Tarazi $^2$ and Cumrun Vafa $^2$}

\affiliation{$^1$Department of Physics, POSTECH, Pohang 790-784, Korea}

\affiliation{$^2$Jefferson Physical Laboratory, Harvard University, Cambridge, MA 02138, USA}

\begin{abstract}
We consider supergravity theories with 16 supercharges in Minkowski space with dimensions $d>3$.  We argue that there is an upper bound on the number of massless modes in such theories depending on $d$.
In particular we show that the rank of the gauge symmetry group $G$ in $d$ dimensions is bounded by $r_G\leq 26-d$.  This in particular demonstrates that 4 dimensional ${\cal N}=4$ SYM  theories with rank bigger than 22, despite being consistent and indeed finite before coupling to gravity, cannot be consistently coupled to ${\cal N}=4$ supergravity in Minkowski space and belong to the swampland.  Our argument is based on the swampland conditions of completeness of spectrum of defects as well as a strong form of the distance conjecture and relies on unitarity as well as supersymmetry of the worldsheet theory of BPS strings.
The results are compatible with known string constructions and provide further evidence for the string lamppost principle (SLP): that string theory lamppost seems to capture {\it  all} consistent quantum gravitational theories.
\end{abstract}

\maketitle

\section{Introduction}

The swampland program \cite{Vafa:2005ui} is based on the assumption that one can draw general lessons about consistency of quantum gravity theories by a careful examination of the string landscape.  If string landscape is only a small fraction of what can be obtained from a consistent theory of quantum gravity, drawing a conclusion based on string theory constructions may be misleading.  It is thus important to evaluate to what extent the string lamppost is complete.  

One of the principles of swampland is the conjecture on boundedness of the number of massless modes in a consistent theory of quantum gravity \cite{Vafa:2005ui}.   This is motivated mainly from the fact that in various string theory constructions it is believed that there are only  a finite number of internal compactifications consistent with a particular amount of supersymmetry.  For example, it is believed that there is a finite number of Calabi-Yau manifolds in each dimension \cite{Yau:1991}.  However could it be that we are being misled by the string lamppost?  Could it be that only the ones we can obtain as part of the string landscape are bounded but there is no intrinsic bound from quantum gravity alone?
To examine this question we consider simplest cases consisting of theories with large amounts of supersymmetry. 

The case of theories with maximal supersymmetry, namely 32 supercharges, is already interesting:  In this case supersymmetry alone fixes the massless matter content of the theory.  So one may think that this case is too boring to check for the string lamppost completeness.  But in fact this is the first example to show that string lamppost is complete:  We can obtain theories with 32 supercharges in all allowed dimensions from string theory.  This did not have to be the case.  The bulk of this paper is devoted to the case with half this maximum amount of supersymmetry in $d>3$ dimensions, namely those with 16 supercharges in $d=4,...,10$.
There are two types of such theories:  Chiral ones which appear only in 10d and 6d, and non-chiral ones which exist in $d=4,...,9$.
The chiral ones are strongly constrained by anomalies.  In particular in $d=10$ the matter can only allow gauge groups $E_{8}\times E_{8}, SO(32), E_8\times U(1)^{248}, U(1)^{496}$ \cite{Green:1984sg}.  However, only the first two are realized in string theory.  One may have thought that the lack of ability to get the latter two cases already demonstrates the inadequacy of the string lamppost, but it has been shown \cite{Adams:2010aa,Kim:2019aa} that these belong to the swampland based on general consistency conditions of supergravity, again reinforcing the completeness of the string theory lamppost.   In $d=6$ the massless spectrum of chiral theories with ${\cal N}=(2,0)$ supersymmetry is completely fixed \cite{Townsend:1983xt} and 
it is indeed the case that this can be constructed in string theory (by compactification of type IIB on $K3$).  This again reinforces the power of the {\it string lamppost principle} (SLP):  that all consistent theories of quantum gravity are in the string landscape.
A similar story repeats for chiral theories with half as much supersymmetry in 6 dimensions with ${\cal N}=(1,0)$ supersymmetry.  In these theories anomaly cancellations lead to infinitely many possible massless spectra \cite{Kumar:2010ru} but only a finite number of them can be constructed using string theory.  Again it was shown that at least a number of the infinite series of anomaly free spectra that have not been constructed in string theory belong to the swampland \cite{Kim:2019aa} (see also \cite{Lee:2019skh}).  This was shown by consideration of unitarity of theories living on the BPS strings in these theories.

A skeptic of the SLP may surmise that the reason for success of SLP in these cases may be because we are dealing only with chiral theories which are strongly constrained by anomalies and if we consider cases without chirality we could find counter-examples to SLP.    It is known from string constructions that for non-chiral theories with 16 supercharges the rank of the gauge group is bounded by $r_G\leq 26-d$\footnote{Note that of course there is nothing inconsistent with a $d$ dimensional 16 supercharge theory of arbitrary rank coupling to gravity in dimensions bigger than $d$.  This is the case for instance for $N$ parallel and coincident D3 branes coupling to $10d$ gravity in a 10-dimensional Minkowski background (leading at large $N$ to the holographic dual $AdS^5\times S^5$).}.  The upper bound on rank can in particular be realized by toroidal compactifications of heterotic strings to lower dimensions.  We will show in this paper that indeed all consistent quantum gravity theories with 16 supercharges must satisfy this bound, reinforcing the SLP. 
The basic idea we use, is to follow the strategy in \cite{Kim:2019aa} and study the theory living on BPS strings in these theories.  We find that supersymmetry and unitarity of QFT on these defects lead to the above bound on the rank.  To show this we need to use in particular a strong form of the distance conjecture \cite{Ooguri:2007aa} (even though this strong version of the distance conjecture is generally assumed it seems it is not explicitly stated except in the recent work \cite{Ooguri_2019,Lee:2019wij,Grimm:2019ixq}).  The strong form of distance conjecture states that at large distances in moduli the light degrees of freedom lead to a dual description of the theory.   Using this we argue that the BPS string wrapped on a circle of radius $R$ should be viewed as a graviton of a dual theory on a circle with radius $1/R$.  This is needed to put bounds on the central charges of the current algebra on the wound string by using the fact that the spin of the dual gravity multiplet cannot be more than 2.

The organization of this paper is as follows:  In section 2 we review some basic structures of theories with 16 supercharges in various dimensions.
We briefly summarize the chiral cases before moving on to review the known constructions for the non-chiral cases.
In section 3 we argue that any consistent quantum gravitational theory with 16 supercharges enjoys T-duality, explaining one of the features observed in string constructions.
In section 4 we study the BPS string in these theories and use supersymmetry combined with unitarity on their worldsheet to place a bound on the rank of the gauge groups for non-chiral theories.  Finally in section 5 we present our conclusion.

\section{Review of supergravity theories with 16 supercharges }
In this section, we would like to review $4\leq d\leq 10$ dimensional theories with 16 supercharges.  Theories with 16 supercharges in $d>3$ consist of  chiral theories in $d=10$ with ${\cal N}=(1,0)$ and $d=6$ with ${\cal N}=(2,0)$ supersymmetry, and non-chiral theories in dimensions $d<10$.  Below we review what is known about these cases.  As we will review below for the chiral theories all the cases allowed by anomaly cancellation and other consistency conditions have already been constructed in string theory.  The non-chiral theories will be the main focus of the present paper.

\subsection{Chiral Theories }

These theories are chiral and are subject to anomaly cancellation conditions which limits the gauge group that appear in 10 and 6 dimensions.

\begin{itemize}
	\item \textbf{$\mathbf{\mathcal{N}=(1,0)}$ in 10d}
\end{itemize}

 In particular, the gauge and gravitational anomaly cancellation in 10d  implemented by the Green-Schwarz mechanism {\cite{Green:1984sg}  requires the gauge groups to be limited to: $E_8 \times E_8 $,$\ SO(32)$, $E_8 \times U(1)^{248},U(1)^{496}$. However, only the first two gauge groups lead to consistent supergravities and the latter two belong to the swampland as argued in \cite{Adams:2010aa, Kim:2019aa}. 

We can construct the two consistent ones in string theory in various ways: $E_8 \times E_8 $, $SO(32)$ heterotic strings, Type I, $SO(32)$ theory and  $E_8\times E_8$ via M-theory on an interval.

\begin{itemize}
	\item 
\textbf{$\mathbf{\mathcal{N}=(2,0)}$ in 6d}
\end{itemize}

These theories have similar anomaly cancellation condition as the 10 dimensional theories. However, vector multiplets are absent in such a theory and hence only the number of tensor multiplet is restricted for a consistent theory coupled to gravity. In particular, a 6d analogue of the Green-Schwarz mechanism \cite{Sagnotti:1992qw} restricts the number of tensor multiplets to be $n=21$ \cite{Townsend:1983xt}.   The moduli space of scalars in this theory is given by \begin{equation}
{SO(21,5)\over SO(21)\times SO(5)} \ .
\label{eq:5} 
\end{equation}
(where here, and in the following we ignore the duality group quotient for the moduli space and all groups are over the reals).

This theory can be understood as the low-energy limit of Type IIB string theory on a $K3$. A T-dual description of this theory can be found by considering M-theory on the orbifold $T^5/\Z_2$ \cite{Witten:1995em,Dasgupta:1995zm}.

\subsection{Non-Chiral Theories }
In this subsection, we review various  known constructions of non-chiral theories  with 16 supercharges in dimensions $3<d<10$ giving more detail for the cases $d=9,8,7$ which have been more thoroughly studied. 
A non-chiral theory with 16-supercharges in $d$ dimensions has matter scalar fields which belong to the coset space 

\begin{equation}
{SO(r_G,10-d)\over SO(r_G)\times SO(10-d)} \ ,
\label{eq:nonchiral} 
\end{equation}
and in addition an $\R^+$ for $d>4$ and a complex axio-dilaton in $d=4$ coming from the gravity multiplet.
Here $r_G$ denotes the rank of the gauge group $G$ and for generic point of the scalar field the $G$ is abelianized to $U(1)^{r_G}$ (In fact as is well known in string theory, different groups $G$ with the same rank can emerge at different points in the vev of scalar fields).
Moreover the gravity multiplets of a non-chiral theory has $(10-d)$ $U(1)$ gauge fields as well as an anti-symmetric two-form gauge field $B_{\mu\nu}$\footnote{Sometimes one considers the dual of this field; for example in 4d the B-field is usually dualized to a scalar axion.}.   We will be interested
in strings charged under the $B$-field later in this paper.

As we will review below, even though classical supergravity alone allows arbitrary rank $r_G$, the ones that appear from string theory constructions all satisfy the bound $r_G\leq 26-d$.    This bound is saturated by considering toroidal compactifications of heterotic strings.
Below we review some of the string theory constructions leading to various lower ranks as well.

\subsubsection{$\mathbf{d=9}$ \textbf{Theories}}

\begin{itemize}
	\item

	{Rank =$17$}

	The 9 dimensional $\mathcal{N}=1$ theory  with  rank 17 can be constructed in many ways.  The simplest way is by considering a circle compactification of heterotic strings.  This can of course be described in many dual ways, including Type  I or M-theory as well. 
	
The moduli space branches for the $E_8 \times E_8$ and $SO(32)$ gauge groups in 9d are actually not distinct but  parts of the full moduli space of rank 17 theories \cite{Narain:1986am}. Each of these components  can be reached by turning on Wilson lines for the heterotic/Type I theories or equivalently  moving D8-branes in Type I'.  Therefore, for rank 17 there exists  one  inequivalent  9 dimensional $\mathcal{N}=1$   theory with moduli space $	SO(17,1)/SO(17)$.

	\item {Rank =$9$}

	There is also one inequivalent  rank $9$ theory which in different regions of its moduli space can be described in terms of   M-theory on a Mobius strip \cite{Park:1998aa,Dabholkar:1996aa},  the  9d CHL string \cite{Chaudhuri:1995aa,Chaudhuri:1995ab} and IIA  with a shift-orientifold  $O8^0$ and $O8^-+7$D8 with an extra D8 \cite{Aharony:2007aa}.
	Hence, we obtain 9 dimensional $\mathcal{N}=1$ theory with gauge group $E_8\times U(1) $ and moduli space given by  $SO(9,1)/SO(9)$.

	\item {Rank =$1$}
	
	There exist two inequivalent rank $1$ theories in 9 dimensions with moduli space given by $SO(1,1)$.
	
	The first one can be obtained from M-theory on the Klein Bottle \cite{Dabholkar:1996aa}  resulting in a theory with 16 supercharges and gauge group $U(1)$. The moduli space of this theory has different weakly coupled descriptions at different regions of the space. In particular, the different regions can be characterized by  M-theory on the Klein Bottle,  IIA with two  shift-orientifolds $O8^0$ and the Asymmetric Orbifold of IIA (AOA)\cite{Aharony:2007aa}.

	The second inequivalent  theory is given  by Type IIB theory on a  Dabholkar-Park(DP) background \cite{Dabholkar:1996aa}.  This latter theory also has various weak coupling descriptions \cite{Aharony:2007aa}.  Upon compactification on a further circle this leads to a T-dual  description of M-theory on Klein Bottle mentioned above.  
Therefore, we have reviewed here two theories in $d=9$ with $r_G=1$ which are distinct, since they have disconnected moduli spaces and can only be connected through T-duality by going to 8d.
\end{itemize}

\subsubsection{$\mathbf{d=8}$ \textbf{Theories}}
Discussions for constructions of theories in $d=8$ can be found  in \cite{deBoer:2001wca, Dabholkar:1996aa} as well as in \cite{Garcia-Etxebarria:2017crf}.

\begin{itemize}
	\item{Rank =$18$}
	
	This theory can be constructed by considering the circle reduction of the 9d rank $17$ theory with an extra $U(1)$ factor  coming from the second circle.

	\item{Rank =$10$}
	
This theory can be viewed as the 8d CHL string \cite{Chaudhuri:1995aa} which is dual to IIA orientifold on the Mobius strip  \cite{Park:1998aa}. The strong coupling limit of the latter description is M-theory on the Mobius strip  which is dual to the 9d CHL string.

	\item{Rank =$2$}
	
There are two inequivalent  9d rank $1$ theories which describe the same theory in  8 dimensions \cite{Aharony:2007aa}.
Therefore, there is one 8 dimensional rank  $2$ theory coming from the 9d  circle reduction with an extra $U(1)$ factor.

\end{itemize}

\subsubsection{$\mathbf{d=7}$ \textbf{Theories}}
The 7 dimensional theories are interesting because they don't all come from a simple  circle compactification of the 8d theories \cite{deBoer:2001wca}. In particular, they find new theories by considering the heterotic string on $T^3$ with some $\Z_n$ triples of commuting holonomies, IIB orientifolds, and F/M-theory compactifications.
\begin{itemize}
	\item {Rank =$19,11$}

These theories are  equivalent to the 8d rank $18,10$ theories compactified on a circle respectively.

\item {Rank =$7$}

There is one theory with rank 7 obtained from the heterotic string on $T^3$ with some $\Z_3$ triples. This theory is dual to F-theory on $K3\times S^1 \over \Z_3$.

\item {Rank =$5$}

Similarly, this theory can be constructed from the heterotic string with $\Z_4$ triples and is dual to F-theory on $K3\times S^1 \over \Z_4$.

	\item{Rank =$3$}
	
	There are in total 4 inequivalent theories with rank $3$.
	In particular, there are two inequivalent theories obtained from the heterotic string with $\Z_{5,6}$ triples and are dual to F-theory on $K3\times S^1 \over \Z_{5,6}$ respectively.
		In addition, there are two inequivalent theories coming from IIA orientifolds $ 4O6^-+4 O6^+$.   They could possibly both be described in M-theory on $K3$ with frozen singularities (assuming two non-isomorphic embeddings of the $(D_4)^4$ weight lattice into the K3 lattice exist) with   dual  F-theory compactification on  ${(T^4\times S^1)\over \Z_2}$.
		\item{Rank =$1$}

Finally, there are three inequivalent theories coming from F-theory on $T^4\times S^1 \over \Z_{3,4,6}$. These three theories do not have a heterotic description but have an M-theory description in terms of $K3$ compactification with frozen singularities.

\end{itemize}

\section{T-duality and the strong version of the distance conjecture}

In this section we use the strong version of the distance conjecture \cite{Ooguri:2007aa} to argue that if we take a non-chiral theory with 16 supercharges in $d$ dimensions and put it on a cirlce of radius $R$ with no additional Wilson lines turned on, in the limit of $ R\rightarrow 0$ it is equivalent to another theory with 16 supercharges again in $d$ dimensions.  Moreover, the winding string of one is equivalent to momentum modes of the dual theory.
 The resulting theory may or may not be the original theory.  An example of this, when you get the same theory back, is compactification of the heterotic string on a circle as the heterotic circle is self-dual under T-duality.  An example of the T-duality which is not a self-duality is the two inequivalent rank 1 theories we discussed in 9 dimensions, which are T-dual of one another.  We will now argue, regardless of the origin of the non-chiral theory, the strong version of the distance conjecture leads to the prediction of T-duality.
As we will see the main ingredient in this argument is that such a large amount of supersymmetry strongly constrains the moduli space of the theory upon circle compactification.

We consider a  $d$-dimensional non-chiral theory with 16 supercharges with a gauge group of rank $r_G$ in the low energy effective action on a circle $S^1$. The moduli space of this theory was written in (\ref{eq:nonchiral}) and upon the circle compactification and in the absence of Wilson lines on the circle it only acquires an extra factor $SO(1,1)$ coming from the radius of the circle. This subspace of the moduli space associated with the vector multiplets is then given by 

\begin{equation}
{\frac{SO(r_G,10-d)}{SO(r_G)\times SO(10-d)}}\times SO(1,1) \ ,
\label{eq:5} 
\end{equation}   

We note that if we were to turn on Wilson lines  for the gauge fields around the circle then some scalars would acquire a vev and the moduli space would become\footnote{We are ignoring the fact that in 3d the gauge bosons can be dualized to scalars and add to the dimension of moduli space.}

\begin{equation}
{\frac{SO(r_G+1,10-d+1)}{SO(r_G+1)\times SO(10-d+1)}} \ ,
\label{eq:5} 
\end{equation}   

However, in the cases we are considering such vevs are not turned on and hence the moduli remains the same when we put on a circle up to the extra $SO(1,1)$.   Note that as already mentioned the gravity multiplet includes a $B_{\mu \nu}$ field.  Compactifying the theory on a circle will lead to a gauge field $B_{\mu \theta}$ where $\theta$ is along the direction of the circle.  Strings charged under $B$ and wound around the circle will continue to carry the charge of the gauge field in the lower dimension.  In addition, there is a gauge field associated to the reduction of the gravity mode on the circle $g_{\mu \theta}$ and the modes carrying $KK$ momentum are charged under this mode.  The supersymmetry algebra implies that BPS states charged under the momentum and winding modes have central charges $(n/R,mR)$ where $(n,m)$ denote the momentum and winding numbers.  

We now consider the circle becoming very small.  Then one would expect the dual theory to at least have $(d-1)$ dimensional Lorentzian symmetry.   We will now argue that the theory should actually end up being $d$ dimensional in the limit $R\rightarrow 0$.
The general statement of the distance conjecture \cite{Ooguri:2007aa} is that at infinite distance in moduli space an infinite tower of light states emerges. A stronger version of this conjecture  states that when this tower appears then there exists a dual weakly-coupled description with its basic modes comprised of the states in the light tower.  In the case at hand as $R\rightarrow 0$ the supersymmetry algebra implies that the winding modes are getting light.  This winding modes should be the dual description of some elementary excitations of a dual theory.  On the other hand
 we have found that the moduli space of the theory whose masses do not depend on $R$ includes
\begin{equation}
{\frac{SO(r_G,10-d)}{SO(r_G)}} \ ,
\label{eq:7} 
\end{equation}
which predicts that the theory must have at least $d$ dimensions, because of the classification of scalar moduli space of theories with 16 supercharges.  Thus the light modes which used to be winding modes of the original theory, must now be part of the weak coupling limits of this d-dimensional theory.  In other words, they must be the momentum modes of this theory.  Said differently we have argued that every non-chiral theory enjoys T-duality.   Of course this argument does not predict whether the T-dual theory is different or the same as the original theory, compatible with the fact that both versions do occur in string theory.
Note that this implies that the singly wound string of the original theory in its ground states should carry the same quantum numbers as a graviton multiplet with one unit of momentum around the circle, since it is dual to it.  In particular it is a massive state in a $d-1$ dimensional theory with maximum spin 2.  We will use this fact below.

\section{Derivation of the bound on the rank}

As we have discussed the supergravity multiplet for non-chiral theories with 16 supercharges includes an anti-symmetric 2-form tensor field $B$.
By completeness of the spectrum we can consider a BPS string charged under this field and study the consequences of supersymmetry and unitary of the worldsheet CFT on this string. From this, we will show that the rank of the gauge group is bounded from above. The strategy to find the bound is exactly the one employed in \cite{Kim:2019aa}.  

The string with tensor charge $Q$ couples to the bulk theory through the following term:
\begin{equation}\label{eq:string-S}
	S^{\rm str} = Q\int_{\mathcal{M}_d}B \wedge \prod_{a=1}^{d-2}\delta(x^a)dx^a = Q\int_{\mathcal{M}_2}B \ ,
\end{equation}
where we assume the string is located at the origin $x^a=0$ of the transverse $\mathbb{R}^{d-2}$ directions.

The 1/2 BPS string will preserve ${\cal N}=(0,8)$ supersymmetry. This can be explained as follows. The supersymmetry algebra in the presence of the string should be comprised only of unbroken symmetry generators. In particular, the momentum generators along the transverse $\mathbb{R}^{d-2}$ cannot be parts of the algebra. First consider the 4d case. Suppose a string is stretched along $x^0,x^3$ directions in 4d supergravity.  
The 4d SUSY algebra involves $\{ Q_\alpha^I, \bar{Q}_{\dot{\beta}J}\}\sim (\sigma^\mu P_\mu)_{\alpha\dot\beta}\delta^I_J$ with the Pauli matrices $\sigma^\mu=(-\mathbf{1},\sigma^i)$. Then for the absence of $P_1,P_2$ generators in this algebra the string will pick  unbroken 8 supercharges $Q^I_+,\bar{Q}_{\dot+ J}$ (or $Q^I_-,\bar{Q}_{\dot- J}$) whose currents turn out to be right-moving (or left-moving) in the 2d worldsheet. Also, half the supercharges in $d>4$ supergravity preserved in the presence of 2d defects reduce to these chiral supercharges under toroidal compactification to 4d (with the string worldsheet transverse to the circles). This shows that the 8 supercharges have a definite chirality on the worldsheet for all $d$. They are right-moving in our convention.

We note that the $B$ field transforms non-trivially under the local Lorentz and the gauge transformations with parameters $\Theta$ and $\Lambda_i$ respectively:
\begin{equation}\label{eq:B-variation}
	\delta B =-\frac{1}{4}{\rm Tr}(\Lambda\cdot F) + \kappa \,{\rm tr}(\Theta R) \ ,
\end{equation}
with the gauge field strength $F_i$'s and the curvature 2-form $R$ in the bulk supergravity.
We have used $a\cdot b=\Omega^{ij}a_ib_j$ for the dot product of two vectors in the charge lattice $\Gamma_{r_G,10-d}$ with respect to the $SO(r_G,10-d)$-invariant metric $\Omega^{ij}$ with signture $(r_G,10-d)$.
Here, the gauge variations of $B$ are fixed by  invariance of the action under 16 supercharges \cite{Bergshoeff:1981um,Chapline:1982ww,Bergshoeff:1985ms}. On the other hand the variation under the local Lorentz transformation is from higher derivative corrections that cannot be fixed solely by supersymmetry. So the coefficient $\kappa$ is yet to be determined by other means.
The bulk action now includes the string action $S^{\rm str}$ that is not invariant under the symmetry transformations due to the variation rules (\ref{eq:B-variation}) of $B$. Therefore, the presence of the string induces anomaly inflow toward the string worldsheet. This anomaly inflow must be cancelled by the anomaly coming from the worldsheet degrees of freedom.

The chiral degrees of freedom on the worldsheet CFT could have non-trivial anomaly and we expect that this worldsheet anomaly cancels the anomaly inflow from the bulk gravity theory discussed above. The cancellation of the anomaly inflow then restricts the anomaly polynomial of the worldsheet CFT on $Q$ strings to the form,
\begin{eqnarray}\label{eq:anomaly-Q}
	I_4 &=& Q\left[-\kappa\,{\rm tr}(R^2)+\frac{1}{4}{\rm Tr}(F\cdot F)\right] \nonumber \\
	 &=& Q\left[\frac{\kappa}{2}p_1(T_2)-\kappa\,c_2(SO(d\!-\!2))+\frac{1}{4}{\rm Tr}(F\cdot F)\right] \ ,
\end{eqnarray}
where $p_1(T_2)$ is the first Pontryagin class of the tangent bundle $T_2$ on the 2d worldsheet and $c_2(SO(d\!-\!2))$ is the 2nd Chern-class of the $SO(d\!-\!2)$ normal bundle for the transverse $\mathbb{R}^{d\!-\!2}$ rotation. Here we used the decomposition ${\rm tr}R^2 = -\frac{1}{2}p_1(T_2)+c_2(SO(d\!-\!2))$. 

When $d=10$, for example, the anomaly polynomial with $\kappa=1$ coincides with that of the 2d CFT on BPS strings in the 10d $\mathcal{N}=(1,0)$ supergravity with $E_8\times E_8$ or $SO(32)$ gauge group computed in \cite{Kim:2019aa}. In this case the constant $\kappa=1$ is fixed by the bulk anomaly cancellation by Green-Schwarz mechanism.

The worldsheet theory at low-energy reduces to a 2d conformal theory with at least $(0,8)$ supersymmetry.  It is a priori conceivable that the supersymmetry gets enhanced in the IR to $(8,8)$.  It is important to distinguish these two possibilities because the anomaly coefficients compute the left minus right contributions.  Let us first discuss the case where there are no enhancements and the theory in the IR has $(0,8)$ supersymmetry only.

For the $(0,8)$ case we can easily compute the central charges of the 2d worldsheet theory from the anomaly polynomial.
First, the coefficient of the gravitational anomaly $-\frac{1}{24}p_1(T_2)$ encodes the relative central charge $c_R-c_L$. In addition, we can obtain the right-moving central charge $c_R$ using the $\mathcal{N}=(0,2)$ superconformal subalgebra in the $\mathcal{N}=(0,8)$ supersymmetric theory. The $U(1)_R$ R-symmetry group of the $\mathcal{N}=(0,2)$ superconformal algebra is chosen as an $SO(2)$ subgroup of the $SO(d-2)$ rotation group (which can be done since $d\geq 4$). Then the $(0,2)$ algebra relates the right-moving central charge with the 't Hooft anomaly $k_R$ of the $U(1)_R$ symmetry such as $c_R=3k_R=12\kappa$ where $\kappa$ is the 't Hooft anomaly coefficient for the $J_{SO(2)}$ current of $SO(2)\subset SO(d-2)$ bulk symmetry related to the $U(1)_R$ current by $J_R= 2J_{SO(2)}$.
As a consequence, we find the central charges,
\begin{equation}\label{eq:centralC}
	c_R = 12 \kappa \ , \quad c_L = 24 \kappa \ .
\end{equation}
$\kappa$ is quantized to be an integer because the $SO(2)\subset SO(d\!-\!2)$ is part of the Lorentz symmetry.

We note that this result involves the contributions from the center-of-mass degrees of freedom that come from the zero modes of broken symmetries in the presence of BPS strings. The center-of-mass modes form a free $(0,8)$ multiplet $(\sigma^a,Y^i,\lambda^+)$ with $a=1,\cdots,10\!-\!d$ and $i=1,\cdots d-2$ where $\sigma^a$ are right-moving compact scalars and $Y^i$ are non-compact scalars (which realize the symmetry currents associated with the $10-d$ graviphotons), and $\lambda^+$ are 8 right-moving fermions. A simple counting yields their central charges $c_R^{\rm com}=12,c_L^{\rm com}=d-2$. Thus the central charges of the interacting sector in the worldsheet theory are
\begin{equation}
	\tilde{c}_R = c_R - c_R^{\rm com} = 12(\kappa-1) \ , \quad \tilde{c}_L = c_L - c_L^{\rm com} = 24\kappa +2-d \ .
\end{equation}

The 't Hooft anomaly coefficients of flavor symmetry groups are identified with the levels of the Kac-Moody current algebra. For a string with $Q=1$ the anomaly polynomial in (\ref{eq:anomaly-Q}) tells us that the 't Hooft anomalies $k_i$ for flavor symmetry groups, which originate from the bulk gauge symmetries, are given by eigenvalues of the metric $\Omega^{ij}$. In our convention, right-moving (or left-moving) current algebra provides negative (or positive) contribution to the associated 't Hooft anomaly. The $k_i$ is a net contribution from both sectors. Therefore one can deduce that the 2d CFT on a single string must contain at least one current algebra for every symmetry group $G_i$ realized in the right-moving sector if $k_i<0$ or in the left-moving sector if $k_i>0$.

Let us now turn to unitarity constraints on the worldsheet of $(0,8)$ CFTs of BPS strings. First, the current algebra of simple non-Abelin group $G$ with level $k$ in the 2d CFT contributes to the central charge by the factor (see, e.g., \cite{DiFrancesco:1997nk})
\begin{equation}
	c_G=\frac{k \, {\rm dim}G}{k + h^\vee} \ ,
\end{equation}
where dim$\,G$ is the dimension and $h^\vee$ is the dual-Coxeter number of group $G$ respectively. For a $U(1)$ current algebra, the central charge contribution is $c_{U(1)}=1$. This leads to the following constraint on the left-moving central charge \cite{Kim:2019aa}:
\begin{equation}\label{eq:condition-on-C}
	\sum_i c_{G_i}  \le \tilde{c}_L \ ,
\end{equation}
where the sum for $i$ is taken over the left-moving currents. 
Remarkably, this inequality provides a strict upper bound on the rank of the bulk gauge group.
On the Coulomb branch where the bulk gauge symmetry is broken to Abelian groups including $U(1)^{r_G}$, we can further simplify the bound as
\begin{equation}
\label{eq:bound} 
	r_G \le \tilde{c}_L = 24\kappa+2-d \ .
\end{equation}
If this bound is violated, the anomaly inflow from the bulk gravity theory cannot be cancelled by the anomalies of a unitary CFT on the string. Hence, a consistent bulk gravity theory involving BPS strings must satisfy this bound.

We now argue that any gravity theory with 16 supercharges that has a well-defined T-dual theory necessairly has $\kappa<2$. For this we consider a string with $Q=1$ wrapped around a circle of radius $R$ and study its ground states.
Consider first the ground states of the wound string in the Ramond sector of the interacting CFT.  As has been argued in \cite{Lerche:1989uy} the maximum charge for the $U(1)_R$ spectrum in the Ramond sector for this theory is given by $c_R/6$ the central charge of the current, i.e. $2(\kappa-1)$ after removing the center-of-mass contribution, which means that the spin is less than or equal to $(\kappa -1)$.    Moreover as argued there this maximum range is actually realized by the spectral flow of the vacuum state of the NS sector to the R sector\footnote{The argument in \cite{Lerche:1989uy} is mainly in the context of (2,2) supersymmetric theory, but that can be easily adapted to the $(0,2)\subset (0,8)$ being considered here, using the integrality of charges of the $U(1)_R$, which follows from the fact that it is twice the spins in physical space which are integer or half-integer.}.

Now we use the fact that the spectrum of the singly wound string already includes spin 2 states arising from the center-of-mass degrees of freedom in the right-moving sector. As we argued, in addition, the internal degrees of freedom from the interacting CFT contains right-movers carrying charges under the $SO(2)\subset SO(d\!-\!2)$ rotational symmetry. This simply means that the ground states of the wound string will include a state with a net spin bigger than 2 when $\kappa-1>0$. This higher spin state is generated by a tensor product of the spin 2 states in the center-of-mass spectrum and the internal right-moving state carrying $SO(2)$ Lorentz charge $\kappa-1$.
On the other hand, a conseqence of T-duality is that the ground states of the wound string should have the same quantum numbers as the gravity multiplet in the dual theory. So T-duality cannot hold if $\kappa>1$ due to the higher spin states. Therefore, we conclude there are only two possibilities
\begin{equation}
	\kappa = 1 \ {\rm or} \ 0 \ .
\end{equation}

When $\kappa=1$, the central charges of the 2d CFT on a single string are $(c_L,c_R)=(24,12)$. The unitarity of this string probe when coupled to the bulk gravity imposes a novel constraint on the rank of the bulk gauge groups,
\begin{equation}
\label{eq:upperbound} 
	r_G \le 26-d \ .
\end{equation}

The case $\kappa =0$ is not allowed for the $(0,8)$ case as can be seen from the fact that there are always the center of mass modes and thus we cannot have $c_L=c_R=0$ which would be a consequence of  (\ref{eq:centralC}).  However, the $\kappa=0$ case is in fact forced on us, as we will  discuss next, for the case when the supersymmetry is enhanced in the IR to $(8,8)$.

We now turn to the case with $(8,8)$ supersymmetry.   In this case we still have 
$$ c_L-c_R=12 \kappa \ ,$$
but in addition we have two $U(1)_R$ currents, one left- and one right-moving with anomaly coefficients $k_L,k_R$, satisfying
$$k_L-k_R=4\kappa\ ,$$
with $c_L=3k_L,c_R=3k_R$.
    By the T-duality argument we just used which implies that we should have no additional spins other than those coming from the center of mass, we learn that $k_L=k_R=1$ and $\kappa =0$.  Moreover, this implies that $c_L=c_R=12$ and subtracting the degrees of freedom coming from the center of mass, we get the bound on the rank of the gauge group $r_G \le 10-d$.  This bound is stronger than the one coming for theories with $(0,8)$ supersymmetry (\ref{eq:upperbound}).  Taking this into account we learn that
any gravity theory with 16 supersymmetries while having higher rank gauge groups beyond the bound in (\ref{eq:upperbound}) are inconsistent and therefore belong to the swampland.

Both the $(0,8)$ and $(8,8)$ supersymmetric cases  on BPS strings are realized in string theory.  In particular, the toroidal compactifications of heterotic string realizes the $(0,8)$ case, consistent with the fact that it has $(c_L,c_R)=(24,12)$.    An interesting example of this is the rank 9 theory in $d=9$.
In this case we can have at the CHL point the $E_8\times U(1)$ matter gauge symmetry.  The central charge of the left-moving degrees of freedom comes from the center of mass contribution of ($c_1=7$) plus $E_8$ at level $k=2$ ($c_2=15 {1\over 2}$) plus the $U(1)$ ($c_3=1$) and a left-over piece which is an Ising model ($c_4={1\over 2}$) leading to $c_L=24$.

The case with $(8,8)$ is also realized:
One such example is the IIA limit of M-theory compactified on Klein bottle, called the AOA theory, introduced in \cite{Aharony:2007aa}. A BPS string in the AOA theory enjoys the $(8,8)$ symmetry enhancement. The 2d theory on a single string consists of bosonic fields $(A_0,A_1,A_2,Y^i)$ with $i=1,\cdots,7$ and a pair of fermionic fields $\psi^\pm$ where $A_0,A_1$ are 2d gauge fields and $A_2$ is a compact (but non-chiral) scalar and $Y_i$ are non-compact scalars, and $\pm$ denotes the 2d chirality \cite{Aharony:2007aa}. The $SO(7)$ Lorentz symmetry acts on both (anti-)chiral fermions $\psi^\pm$ as well as $Y_i$. These fields form a free $(8,8)$ multiplet for the center-of-mass degrees of freedom of the string. From the matter content, one can read off the central charges as $c_R=c_L=12$.  We also notice that the 't Hooft anomalies for the $U(1)\times U(1)$ gauge symmetry receive contributions  $+1$ and $-1$, respectively, from the left- and the right-moving components of the compact scalar $A_2$, which precisely cancel the anomaly inflow for $U(1)\times U(1)$ gauge symmetry.
Toroidal compactifications of this theory give rise to other examples with $\mathcal{N}=(8,8)$ enhanced supersymmetry in $\kappa=0$ gravity theories.

Note that the upper bound on the rank 
(\ref{eq:upperbound}) is saturated by the toroidal compactifications of 10d heterotic strings to lower dimensions. By turning on holonomies in the compactifications, we can also construct a large class of lower rank theories. Hence the above rank bound provides a strong evidence for the string lamppost principle.

\section{Conclusion} 
We have seen that the number of massless modes in a supersymmetric theory of gravity with 16 supercharges admitting Minkowski background is bounded.  In particular the massless matter in such theories which is determined by the choice of a gauge group $G$, satisfies a bound on its rank:
$r_G\leq 26-d$.  Moreover as we have seen this bound is satisfied by all the known string theory constructions. 

There are a number of ways one may hope to extend the results in this paper:  One direction is to find a further refinement on the ranks.  For example the string constructions suggest that in $d=9$ the only allowed ranks are $17,9,1$.  There are similar restrictions from known string constructions on the actual ranks that do appear in other dimensions.  Can one derive these bounds as well?  Another restriction is to find the actual gauge groups that can appear in consistent theories, which are compatible with the rank conditions.  It is known that not all groups which are consistent with the rank condition seem to arise.   For example in $d=8$ some gauge groups are forbidden by global anomalies \cite{Garcia-Etxebarria:2017crf}.   However, even taking into account global anomalies some cases, like $G_2$ gauge group does not seem to appear as part of consistent theory with 16 supercharges in $d=8$.
It would be interersting to find all the allowed groups that can appear.

This paper is part of the larger program of finding an upper bound on the number of massless modes in supersymmetric gravity theories.
We now know that theories with $\mathcal{N}=32,16$ supercharges have a fixed upper bound on the number of massless modes in each dimension which are indeed all realized by string constructions.  It would be natural to continue this to theories with less supersymmetry, and in particular to the next case with $\mathcal{N}=8$ supercharge theories in 6 and lower dimensions and check whether the SLP still continues to hold.

\textbf{Acknowledgments} 

We would like to thank N. Berkovits, N. Bobev, Z. Komargodski, M. Montero, M. Rocek and F. Toppan for valuable discussions. HK would like to thank Harvard University for hospitality during part of this work.

The research of HK is supported by the POSCO Science Fellowship of POSCO TJ Park Foundation and the National Research Foundation of Korea (NRF) Grant 2018R1D1A1B07042934. 
The research of HCT and CV is supported in part by the NSF grant PHY-1719924 and by a grant
from the Simons Foundation (602883, CV).


\let\bbb\bibitem\def\bibitem{\itemsep4pt\bbb}
\bibliography{ref}

\end{document}